\documentclass[aps,prl,twocolumn,superscriptaddress,floatfix]{revtex4-2}

\usepackage{graphicx}
\usepackage{amsmath}
\usepackage{amssymb}
\usepackage{hyperref}
\usepackage{color}
\usepackage{orcidlink}

\usepackage{xcolor}
\definecolor{myblue}{RGB}{0,70,140}
\hypersetup{
    colorlinks=true,
    allcolors=myblue
}

\begin{document}

\title{Uniform-Charge-Density $d$-Wave Superconductivity in the Pure $t$-$J$ Model at $1/8$ Doping on an Infinite Cylinder}

\author{Guangyu Yu\orcidlink{0009-0000-7405-1853}}
\affiliation{Kavli Institute for Theoretical Sciences and School of Quantum, University of Chinese Academy of Sciences, Beijing 100190, China}

\author{Zheng Zhu\orcidlink{0000-0001-7510-9949}}
\email{zhuzheng@ucas.ac.cn}
\affiliation{Kavli Institute for Theoretical Sciences and School of Quantum, University of Chinese Academy of Sciences, Beijing 100190, China}

\date{\today}

\begin{abstract}
The ground state of the square-lattice  $t$-$J$ model at $1/8$ hole doping has been the subject of a long-standing controversy, with conflicting numerical reports of stripe order versus superconductivity. 
Here, we address this debate using the variational uniform matrix product state (VUMPS) method, which
optimizes the wave function directly in the thermodynamic limit along the cylinder axis and eliminates the boundary pinning effects that can arise in finite-cylinder calculations.  On an infinite cylinder of circumference $L_y = 8$, we find a uniform-charge-density $d$-wave superconducting ground state with quasi-long-range pairing correlations. The dominant pairing occurs at zero center-of-mass momentum, while no sizable finite-momentum pairing component is detected.
Twisted-boundary-condition calculations reveal a finite superconducting phase stiffness, with the order-parameter phase winding smoothly to follow the applied flux, providing an independent probe of the robustness of the superconductivity. Moreover, both charge and spin correlations decay rapidly, with no signatures of long-range stripe or magnetic order, in contrast to an $L_y=6$ cylinder at $1/6$ doping where the same method readily identifies stripe order.
Our results provide methodologically distinct evidence in the long-standing  stripe--superconductivity debate in the pure $t$-$J$ model and reveal   uniform $d$-wave superconductivity in a minimal model of doped Mott insulators.  

\end{abstract}

\maketitle

\textbf{Introduction.—}
Since the discovery of high-temperature superconductivity (HTS) in cuprates~\cite{Bednorz1986},
identifying the minimal theoretical model that captures the essential physics of HTS has remained a central challenge in condensed matter physics~\cite{Imada1998,Orenstein2000,Lee2006,Armitage2010,Keimer2015}. The Hubbard and $t$-$J$ models are the most prominent candidates~\cite{ZhangRice1988,Anderson2004,Arovas2022,Qin2022Review}, but inconsistent numerical results for both have obscured the path to a consensus~\cite{Lee2006,Ogata2008}.

The $t$-$J$ model is widely regarded as the effective low-energy model of the Hubbard model in the strong-coupling limit,
where double electron occupancy is strictly excluded. One might expect the two models to be qualitatively equivalent. However, accumulating numerical evidence contradicts this view, particularly at $1/8$ hole doping, a concentration where the superconducting transition temperature $T_c$ is strongly suppressed in cuprates~\cite{Moodenbaugh1988}.
For the  square-lattice Hubbard model at $1/8$ doping, most numerical studies report a stripe order phase 
as the ground state
~\cite{Zheng2017,Huang2018,Jiang2019Science,Qin2020,Chung2020,Jiang2020PRR,Jiang2024,Xu2024,Jiang2026Competition,r4q9-4yvj,Gu2026Solving}. For the $t$-$J$ model,  however, the nature of the ground state at $1/8$ doping  has remained under debate for more than two decades, with competing scenarios involving stripe order and superconductivity~\cite{White1998,PhysRevB.60.R753,Sorella2002,Raczkowski2007,Corboz2011,Hu2012,Corboz2014,Jiang2018,PhysRevLett.132.066002,Chen2025PNAS}. Related studies of extended and frustrated $t$-$J$ models have further revealed a rich competition among superconductivity, stripe order, and intertwined phases~\cite{Capello2008,Dodaro2017,Li2021iPEPS,Jiang2021PNAS,Jiang2022tttJ,Gong2021,Jiang2023,Chen2024,PhysRevB.110.165127,2f9b-h135,Blatzshort18Leg2026,Wang2026}.
The persistence of this disagreement across successive generations of numerical methods suggests that the reported ground state may depend sensitively on details of the simulation setup, such as the open boundaries and finite lengths of conventional finite-cylinder calculations. This possibility calls for an assessment with an independent numerical method.

Here we revisit this long-standing issue from a methodologically different angle, employing the variational uniform matrix product state (VUMPS) method~\cite{Zauner-Stauber2018,Vanderstraeten2019} on an infinite cylinder. Unlike finite-system simulations, VUMPS optimizes the ground-state wave function directly in the thermodynamic limit along the cylinder axis, eliminating open boundaries and finite-length effects in this direction and thereby providing an independent assessment of the competition between stripe order and superconductivity. VUMPS can also achieve well-converged observables at moderate bond dimensions, reducing the computational effort required to access the infinite-cylinder regime. This method has demonstrated its reliability in various strongly
correlated and topological systems,
including quantum spin systems~\cite{Rausch2025Sawtooth,Fuji2026SpinChains}, the half-filled Hubbard model~\cite{VanDamme2021}, quantum Hall bilayers~\cite{Yu2025QHBilayer}, and fractional Chern
insulators~\cite{yu2026compositebosontheoryfractional}.

On the $L_y=8$ infinite cylinder,
we find that the ground state of the pure $t$-$J$ model at $1/8$ doping is a robust uniform-charge-density $d$-wave superconductor with quasi-long-range order and no detectable long-range charge or magnetic order.  A weak period-two modulation of the bond order parameter is present at finite bond dimension, but its extrapolated weight relative to the uniform component decreases rapidly with increasing $\chi$; together with a pair structure factor that peaks only at zero momentum, this indicates the absence of a sizable pair-density-wave (PDW) component.  
We further corroborate the robustness of the superconducting state via twisted boundary condition simulations: the ground-state energy acquires a quadratic stiffness cost, and the order-parameter phase winds smoothly to follow the applied flux even for twists as large as $60^\circ$. 
Importantly, the absence of stripe signatures is not a generic consequence of the infinite-cylinder ansatz: the same VUMPS setup readily detects stripe order on an $L_y = 6$ cylinder at $1/6$ doping.
Our results thus add a methodologically distinct piece of evidence to this decades-old debate, bearing directly on the status of the $t$-$J$ model as a minimal model for HTS.

\textbf{Model and method.—}
The Hamiltonian of the two-dimensional $t$-$J$ model is
\begin{equation}
H=-t\sum_{\langle ij\rangle, \sigma}\left(c^\dagger_{i\sigma} c_{j\sigma} + \text{h.c.}\right)
+J\sum_{\langle ij\rangle} \left( \mathbf{S}_i \cdot \mathbf{S}_j - \frac{1}{4} n_i n_j \right),
\label{eq:Hamiltonian}
\end{equation}
where $c^\dagger_{i\sigma}$ ($c_{i\sigma}$) creates (annihilates) an electron of spin $\sigma$ at site $i$, $\mathbf{S}_i$ is the spin operator, and $n_i = \sum_\sigma n_{i\sigma}$ is the electron number operator. The first term describes electron hopping with amplitude $t$ (set as the energy unit throughout this work), and the second term is the antiferromagnetic super-exchange interaction with coupling strength $J$. The Hamiltonian acts exclusively on the Hilbert space with no double occupancy. We adopt $J/t = 0.4$, a value widely used for cuprates, and target a hole doping of $1/8$, corresponding to an average electron filling $\langle n_i \rangle = 7/8 = 0.875$.

To accurately determine the ground state, we employ the VUMPS method~\cite{Zauner-Stauber2018,Vanderstraeten2019}  and 
perform simulations on an infinitely long cylinder with circumference $L_y = 8$ along the $y$-direction, using a 16-site unit cell
that accommodates a possible period-2 structure. {Importantly, although expectation values of the optimized state are constrained by this unit cell, possible longer-period ordering tendencies can be examined independently through correlation functions and static structure factors below.}
We impose $U(1)\times U(1)$ symmetry and push bond dimensions up to $\chi = 12000$, for which the relevant bulk
observables are already well converged, together with systematic finite-entanglement scaling toward $\chi\rightarrow\infty$.

\begin{figure*}[tbp]
\centering
\includegraphics[width=0.88\textwidth]{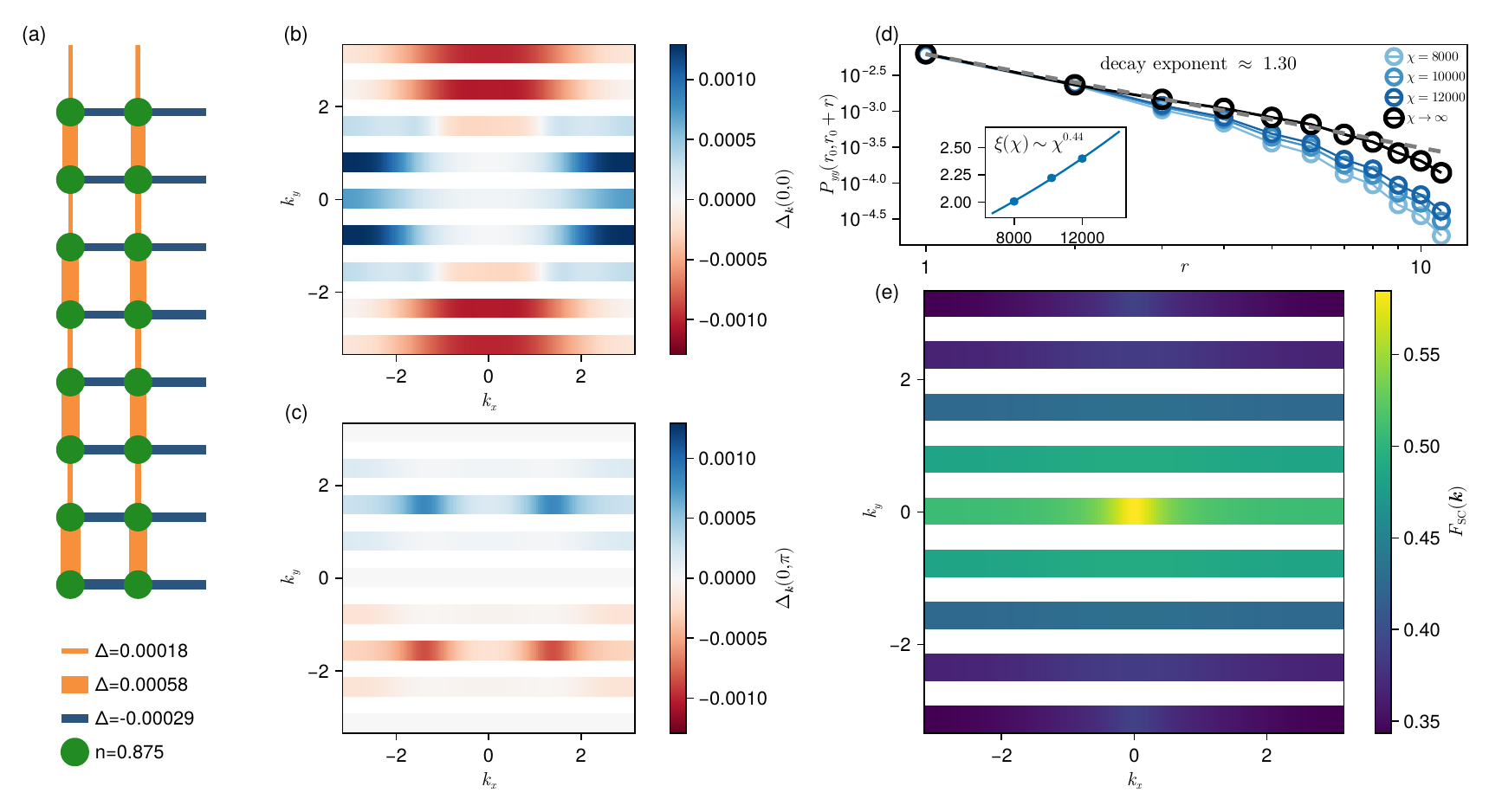}
\caption{(a) Bond-resolved superconducting order parameter and on-site electron density. The two inequivalent $y$ bonds have $\Delta = +0.00018$ (thin orange) and $+0.00058$ (thick orange), while $x$ bonds have $\Delta = -0.00029$ (blue), realizing the $d$-wave sign structure; the density is uniform at $n = 0.875$. (b) Uniform pairing component $\Delta_{\mathbf{k}}(0,0)$ in the first Brillouin zone, shown as discrete strips at the allowed transverse momenta $k_y = n\pi/4$ ($k_x$ continuous); the sign change across $k_x=\pm k_y$ is characteristic of $d_{x^2-y^2}$ pairing.
(c) Finite-momentum component $\Delta_{\mathbf{k}}(0,\pi)$ on the same color scale as (b), strongly suppressed relative to the uniform component.
(d) Singlet pair correlation $P_{yy}(r)$ on a double-logarithmic scale for different bond dimensions and the $\chi\rightarrow\infty$ extrapolation, which follows a power-law decay with $K_{\rm SC}\approx1.30<2$. Inset: pair correlation length, $\xi_{\rm pair}\sim\chi^{0.44}$.
(e) Pair structure factor $F_{\rm SC}(\mathbf{k})$, showing a single dominant
peak at $\mathbf{k}=(0,0)$ and no enhanced finite-momentum weight.
}
\label{fig:order}
\end{figure*}

\textbf{Superconducting pairing.—} We begin by examining the possible superconductivity on the infinite cylinder. The singlet Cooper-pair operator on a bond along direction $\hat{\mathbf{e}}$ is 
\begin{equation}
\hat{\Delta}_{\mathbf{r} \hat{\mathbf{e}}} = \frac{1}{\sqrt{2}}\left(c_{\mathbf{r}\uparrow} c_{\mathbf{r}+\hat{\mathbf{e}}\downarrow} - c_{\mathbf{r}\downarrow} c_{\mathbf{r}+\hat{\mathbf{e}}\uparrow}\right),
\label{eq:pairing}
\end{equation}
and we extract its value on each bond from the leading eigenvector of the MPS transfer matrix in the charge-$2e$ sector. 
As shown in Fig.~\ref{fig:order}(a), the order parameter is spatially uniform on the $x$-oriented bonds, whereas it exhibits a period-two modulation on the  $y$-oriented bonds. The $y$-oriented bonds can be decomposed as $\Delta_y=\overline{\Delta}_y + \delta\Delta_y(-1)^y$ with $\overline{{\Delta}}_y= +0.00038$ and $\delta\Delta_y = 0.00020$. Together with the uniform $x$-bond component, $\Delta_x=-0.00029$, the uniform components are comparable in magnitude and opposite in sign, revealing an underlying $d_{x^2-y^2}$-wave structure. 
The staggered component $\delta\Delta_y$ would correspond to finite-momentum pairing at $Q_y=\pi$; however, finite-entanglement scaling of the bond-resolved pattern shows that it rapidly diminishes with increasing bond dimension, and in the $\chi\to\infty$ limit the extrapolated staggered component amounts to only a few percent of the uniform component. This suggests that the residual modulation is a finite-bond-dimension effect rather than a genuine PDW instability, which is also supported by the momentum-space analysis below.

To further characterize the pairing structure, we resolve the pair wave function in momentum space from 
\begin{equation}
\Delta_{\mathbf{k}}(\mathbf{Q}) = \langle c_{-\mathbf{k}+\mathbf{Q}/2,\downarrow}\, c_{\mathbf{k}+\mathbf{Q}/2,\uparrow} \rangle,
\label{eq:Delta_k}
\end{equation}
where $\mathbf{Q}$ denotes the center-of-mass momentum of the pair and  $\mathbf{k}$ is the relative momentum. 
As shown in Fig.~\ref{fig:order}(b), the uniform component $\Delta_{\mathbf{k}}(0,0)$ is vanishingly small along the diagonal directions $k_x=\pm k_y$, and changes sign across these nodal lines, directly revealing the characteristic of  $d_{x^2-y^2}$-wave pairing. The staggered component identified in real space would correspond to finite-momentum
pairing at $\mathbf{Q}=(0,\pi)$. To assess its relative weight, Fig.~\ref{fig:order}(c) plots
$\Delta_{\mathbf{k}}(0,\pi)$ on the same color scale as the uniform component in Fig.~\ref{fig:order}(b): the finite-momentum component is strongly suppressed relative to the uniform $d$-wave component already at finite bond dimension, consistent with the scaling analysis of the real-space pattern discussed above.

As an independent and more direct diagnostic of the spatial structure of pairing, we compute the pair structure factor, i.e., the Fourier transform of the pair-pair correlation function,
\begin{equation}
F_{\rm SC}(\mathbf{k}) = \frac{1}{N}\sum_{\mathbf{r},\mathbf{r}'} \left\langle \hat{\Delta}^\dagger_{\mathbf{r}\hat{y}}\, \hat{\Delta}_{\mathbf{r}'\hat{y}} \right\rangle e^{i\mathbf{k}\cdot(\mathbf{r}-\mathbf{r}')}.
\label{eq:FSC}
\end{equation}
A PDW would generate a finite-$\mathbf{Q}$ peak in $F_{\rm SC}(\mathbf{k})$, but Fig.~\ref{fig:order}(e) 
instead exhibits a single pronounced peak at $\mathbf{k}=(0,0)$ and no enhanced weight at $(0,\pi)$ or any other finite momentum. 
This demonstrates that the dominant pairing on the $L_y=8$ cylinder is the uniform $Q=0$ component, with no detectable sizable PDW contribution.

\textbf{Pair correlations and phase stiffness.—}
To characterize the long-distance behavior of this superconducting state on the infinite cylinder, we next examine the pair correlations. We define the equal-time correlation between two $y$-oriented singlet pairs separated by a distance $r$ along the infinite ($x$) direction as
$P_{yy}(r)=
\left\langle
\hat{\Delta}^{\dagger}_{\mathbf{r}_0\hat{y}}\,
\hat{\Delta}_{(\mathbf{r}_0+r\hat{x})\hat{y}}
\right\rangle$. 
An important feature of infinite-MPS calculations is that a finite bond dimension $\chi$ necessarily imposes a finite transfer-matrix correlation length. Consequently, even for a state with algebraically decaying correlations, a finite-$\chi$ MPS develops a finite-entanglement cutoff and the correlations eventually cross over to an exponential-like decay at sufficiently long distances~\cite{SCHOLLWOCK201196,Vanderstraeten2019}.   This behavior is clearly visible in the raw $P_{yy}(r)$ data in
Fig.~\ref{fig:order}(d) for finite bond dimensions. We quantify this finite-entanglement cutoff by fitting the long-distance tail of
each finite-$\chi$ correlation function to
$P_{yy}(r)\sim e^{-r/\xi_{\rm pair}(\chi)}$. The resulting effective correlation length grows systematically with bond dimension and is well described by $\xi_{\rm pair}(\chi)\sim\chi^{0.44}$,
as shown in the inset of Fig.~\ref{fig:order}(d). Its continued growth with $\chi$
is consistent with a diverging pair correlation length in the
$\chi\rightarrow\infty$ limit, indicating that the rapid decay of the raw
finite-$\chi$ data is controlled by finite entanglement rather than by an
intrinsically short-ranged pairing state.

To further remove this finite-entanglement cutoff, we extrapolate $P_{yy}(r)$
at each fixed distance to $\chi\rightarrow\infty$ using a second-order
polynomial in $1/\chi$. The extrapolated correlation function exhibits an
algebraic decay, $P_{yy}(r)\sim r^{-K_{\rm SC}}$, with $K_{\rm SC}\approx1.30$. On a quasi-one-dimensional cylinder,
$K_{\rm SC}<2$ implies a divergent superconducting susceptibility
\cite{Kosterlitz1973}.  Importantly, even at the bond dimension $\chi=8000$, the effective power-law exponent in the accessible algebraic-decay regime is already below 2, and the pairing correlations become progressively longer-ranged with increasing $\chi$. Together with the growing finite-entanglement correlation length, these results provide consistent evidence for quasi-long-range superconducting order on the $L_y=8$ infinite cylinder.

We next probe the rigidity of this superconducting state through twisted
boundary conditions along the circumference of the cylinder~\cite{Fisher1989}. A charge flux
$\theta_C$, which couples equally to both spin species, is implemented as
$c^\dagger_{\sigma,L_y+1} \equiv e^{i\theta_C}c^\dagger_{\sigma,1}$
(and similarly across the opposite boundary).
This twist breaks time-reversal symmetry but preserves spin SU(2) symmetry; physically, it corresponds to inserting a magnetic flux through the cylinder,
which couples to the charge sector and probes the superconducting phase
stiffness.
For comparison, we also introduce a spin flux $\theta_S$, which twists the two spin species in opposite directions,
$c^\dagger_{\uparrow,L_y+1} \equiv e^{i\theta_S}c^\dagger_{\uparrow,1}$ and
$c^\dagger_{\downarrow,L_y+1} \equiv e^{-i\theta_S}c^\dagger_{\downarrow,1}$,
preserving time-reversal symmetry but breaking spin SU(2). The corresponding transverse
super-exchange term transforms as $S^+_{1}S^-_{L_y}+\mathrm{h.c.}
\rightarrow
e^{2i\theta_S}S^+_{1}S^-_{L_y}+\mathrm{h.c.}$.

\begin{figure}[tbp]
\centering
\includegraphics[width=\columnwidth]{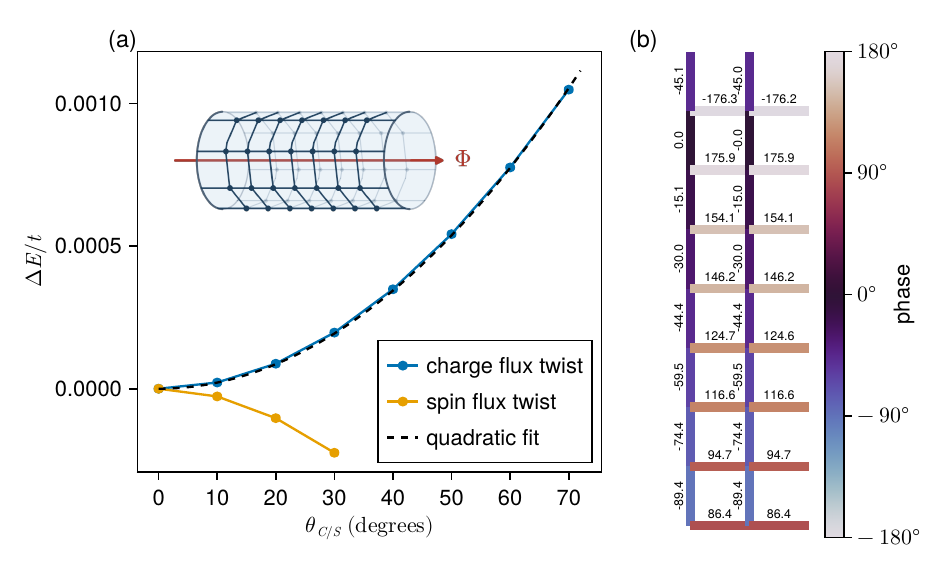}
\caption{(a) Ground-state energy change per site, $\Delta E/t = [E(\theta)-E(0)]/t$, versus twist angle for charge flux (blue) and spin flux (yellow) twists; the inset sketches a flux $\Phi$ threading the cylinder. The energy increases quadratically under charge flux (dashed line: quadratic fit), signaling finite superconducting phase stiffness, but decreases monotonically under spin flux. (b) Order-parameter phase (in degrees, labeled on each bond) under a charge flux twist $\theta_C = 60^\circ$: the phase winds smoothly along the $y$-direction, faithfully following the applied boundary twist.
}
\label{fig:twist}
\end{figure}

As shown in Fig.~\ref{fig:twist}(a), the responses to the two twists are
qualitatively different. Under a charge flux, the ground-state energy develops a positive quadratic curvature around $\theta_C=0$, $E(\theta_C)-E(0)\propto\theta_C^2$,
demonstrating a finite superconducting phase stiffness. 
At the same time, as
directly visualized in Fig.~\ref{fig:twist}(b), the phase of the superconducting
order parameter winds smoothly along the circumference, faithfully following
the applied boundary twist; even at a twist angle as large as
$\theta_C=60^\circ$, the phase profile remains smooth, showing that the system
stays continuously connected to the superconducting state throughout the scan
and attesting to the robustness of the superconducting state. 
In contrast, the spin-flux response does not show a corresponding quadratic
energy increase around $\theta_S=0$. We therefore find
no evidence for an analogous spin-sector rigidity.
Taken together with the
quasi-long-range pair correlations in Fig.~\ref{fig:order}(d), the finite
charge-sector rigidity provides an independent consistency check of the
superconducting character of the ground state on the infinite cylinder.

We note that the charge-flux branch can be followed continuously up to
$\theta_C\approx70^\circ$, while the spin-flux branch shown here extends to
$\theta_S\approx30^\circ$. Beyond these angles, the system evolves into
competing states, and these data are therefore not included in the stiffness
analysis.

\begin{figure}[t]
\centering
\includegraphics[width=\columnwidth]{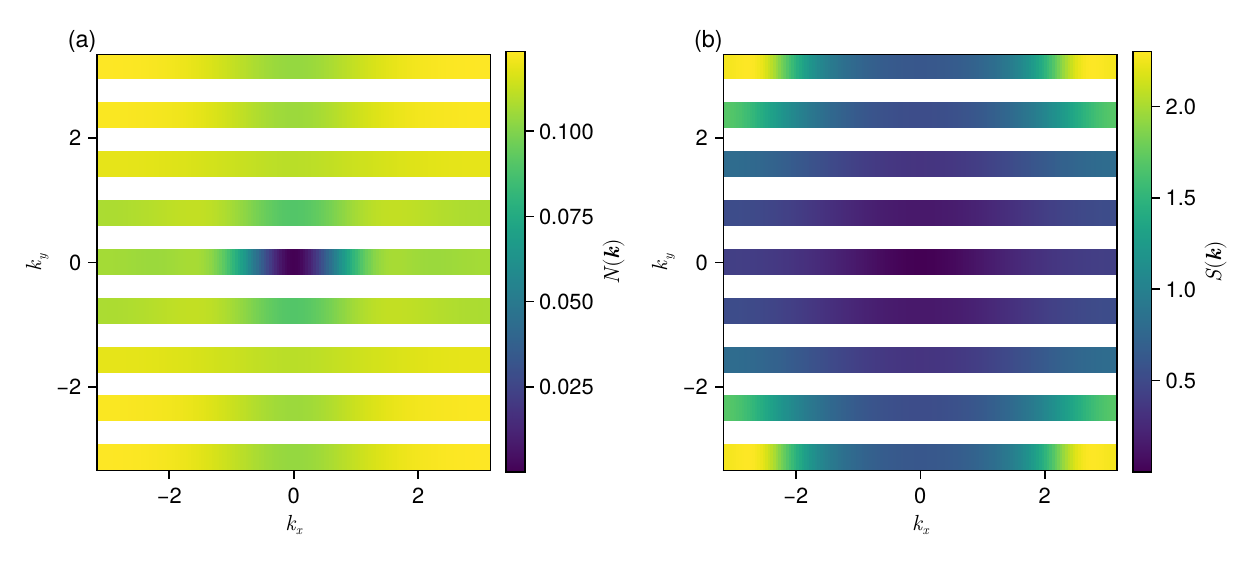}
\caption{(a) Static charge structure factor $N(\mathbf{k})$ in the first Brillouin zone (transverse momentum quantized to the discrete cuts $k_y = n\pi/4$, $k_x$ continuous). The featureless profile shows no evidence
for charge ordering.
(b) Static spin structure factor $S(\mathbf{k})$. The spectral weight near $(\pm\frac{7}{8}\pi, \pi)$ reflects short-range incommensurate antiferromagnetic correlations, while the absence of sharp peaks rules out long-range magnetic order.}
\label{fig:structure}
\end{figure}

 \textbf{Absence of competing orders.—}
We now examine whether the superconductivity  coexists with competing charge or magnetic order. As already seen in Fig.~\ref{fig:order}(a), the on-site electron density is spatially uniform. However, a uniform one-point density does not by itself exclude charge order, since a translationally symmetric state may represent a superposition of symmetry-related charge-ordered configurations. Such hidden ordering tendencies are directly encoded in density correlations. We therefore compute the static charge structure factor,
\begin{equation}
N(\mathbf{k}) = \frac{1}{N} \sum_{\mathbf{r}, \mathbf{r}'} \langle (n_{\mathbf{r}}-\langle {n}_\mathbf{r}\rangle) (n_{\mathbf{r}'}- \langle n_{\mathbf{r}'} \rangle) \rangle\, e^{i\mathbf{k} \cdot (\mathbf{r} - \mathbf{r}')}.
\label{eq:Nk}
\end{equation}
As presented in Fig.~\ref{fig:structure}(a), $N(\mathbf{k})$ is featureless across the Brillouin zone, exhibiting only a smooth, low-intensity profile without any sharp peaks or pronounced modulations, indicating the absence of charge density wave (CDW) order. Although the period-2 unit cell constrains local expectation values, the correlation functions are not restricted to this periodicity. The transfer-matrix spectrum of the infinite MPS retains
sensitivity to fluctuations at arbitrary momenta along the infinite direction. Therefore, any longer-period stripe tendency, such as the period-8 patterns reported in previous finite-system studies, would manifest as enhanced weight in $N(\mathbf{k})$ or long-wavelength oscillations in real-space density correlations. No such signatures are observed. 

Consistently, the real-space density-density connected correlation function $|\langle n_i n_{i+r} \rangle_c|$ decays rapidly with distance and can be fitted either by a steep power law with exponent $\alpha \approx 3.7$ or by an exponential
with a short correlation length $\xi \approx 1.3$ lattice constants (see
Supplemental Material). Both rule out long-range charge order. 
These results provide direct evidence for a uniform charge density in real space and stand in striking contrast to the stripe order found in the Hubbard model at $1/8$ doping. This discrepancy may partly originate from additional terms generated in the
strong-coupling expansion of the Hubbard model, such as the three-site hopping
processes omitted in the pure $t$-$J$ model, which can modify the competition
between superconductivity and charge order~\cite{Yang2024}. 
We stress that the featureless $N(\mathbf{k})$ reported here is not a limitation of the small unit cell: applying the same VUMPS setup to the $L_y=6$ cylinder at $1/6$ doping, where stripe order is well established, resolves sharp charge peaks at the expected wavevector even though the one-point density remains uniform (see Supplemental Material), in sharp contrast to the present case.  

We next examine the magnetic sector through the static spin structure factor
\begin{equation}
S(\mathbf{k}) = \frac{1}{N} \sum_{\mathbf{r}, \mathbf{r}'} e^{i\mathbf{k} \cdot (\mathbf{r} - \mathbf{r}')} \langle \mathbf{S}_{\mathbf{r}} \cdot \mathbf{S}_{\mathbf{r}'} \rangle.
\label{eq:Sk}
\end{equation}
As shown in Fig.~\ref{fig:structure}(b), the  spectral weight is concentrated near $(\pm\frac{7}{8}\pi, \pi)$, consistent with incommensurate antiferromagnetic correlations. The absence of sharp peaks indicates the absence of long-range magnetic order.
Consistently, the real-space spin correlation decays rapidly with distance and can be fitted either by a steep power law with exponent $\alpha_s\approx2.95$ or by an exponential with a short correlation length $\xi_s\approx1.5$ lattice constants (see Supplemental Material). Both rule out long-range magnetic order.

We therefore find a state in which superconducting correlations are
quasi-long-ranged, while both charge and spin correlations decay rapidly,
leaving superconductivity as the dominant long-distance ordering tendency.

 \textbf{Discussion and summary.—}
Our results provide numerical evidence that, on an infinitely long cylinder of circumference $L_y = 8$, the ground state of the pure $t$-$J$ model at $1/8$ doping is a uniform $d$-wave superconductor with quasi-long-range order, and no detectable long-range charge or magnetic order. By eliminating the open boundaries and finite-length effects inherent to finite-cylinder calculations, our infinite-cylinder results provide a methodologically independent reassessment of the long-standing competition between superconductivity and stripe order. This suggests that boundary pinning and finite-length effects may partly account for the differing conclusions reported in the literature.

Two caveats deserve emphasis.
First, while the $L_y=8$ cylinder represents one of the largest circumferences accessible to infinite-cylinder calculations for this model, a single circumference alone does not establish the full two-dimensional limit. Further studies at larger circumferences will be important for determining how the competition between superconductivity and stripe order evolves toward $L_y\rightarrow\infty$. Second, although the period-2 unit cell restricts explicit translational symmetry breaking in the variational ansatz, the correlation functions remain sensitive to longer-period ordering tendencies. The absence of finite-wavevector charge peaks and the rapidly decaying density correlations provide no evidence for an incipient stripe instability on the present cylinder.

Our results also highlight a marked difference between the pure $t$-$J$ model and the Hubbard model at $1/8$ doping. While the Hubbard model stabilizes a stripe phase, on the $L_y = 8$ infinite cylinder the pure $t$-$J$ model instead favors a uniform superconducting state.  
This contrast raises the possibility that additional strong-coupling processes, including three-site hopping terms absent from the pure $t$-$J$ Hamiltonian, may influence the delicate balance between stripe order and superconductivity~\cite{Yang2024, Blatz2025}.

We also clarify the relation to PDW physics, which has been widely discussed theoretically~\cite{Himeda2002,Berg2009,Lee2014,Fradkin2015,PhysRevB.108.115136,Zheng2025,Chen2026PDW,li2026fluctuatingpairdensitywave,Agterberg2020} and reported in cuprate experiments~\cite{Hamidian2016,Edkins2019}. We find no sizable PDW component on the
present geometry. The weak period-two bond modulation
is much smaller relative to the uniform component already at finite bond dimension, rapidly suppressed under finite-entanglement scaling, and leaves no signature in the pair structure factor. Whether PDW correlations emerge at other parameters of the pure $t$-$J$ model remains an  open question.

More broadly, the present work demonstrates that infinite-system tensor-network methods provide a powerful route for resolving closely competing orders in doped Mott insulators. Extensions across doping concentrations, interaction parameters and finite temperatures~\cite{PhysRevX.11.031007,Qu2024,nfmp-32tt,4588-hpc2,Glittum2025,Zhang_2026} will help clarify the evolution of superconductivity and competing
orders in doped Mott insulators and their connection to cuprate phenomenology~\cite{Tranquada1995,Abbamonte2005,Li2007,Ghiringhelli2012,Chang2012}.

\begin{acknowledgments}
We thank Shuai Chen, Alexander Wietek, Yang Qi, Qianqian Chen, Qianghua Wang and Dingping Li for helpful discussions. This work was supported by the National Natural Science Foundation of China (Grant No. 92477106) and the Fundamental Research Funds for the Central Universities.
\end{acknowledgments}

\bibliography{refs}

\newpage
\onecolumngrid
\setcounter{page}{1}
\setcounter{figure}{0}
\setcounter{equation}{0}
\renewcommand{\thefigure}{S\arabic{figure}}
\renewcommand{\theequation}{S\arabic{equation}}
\begin{center}
{\Large \textbf{Supplemental Material for}}\\[0.8em]
{\large Uniform-Charge-Density $d$-Wave Superconductivity in the Pure $t$-$J$ Model at $1/8$ Doping on an Infinite Cylinder}\\[1.2em]
Guangyu Yu and Zheng Zhu\\[0.4em]
{\small Kavli Institute for Theoretical Sciences and School of Quantum, University of Chinese Academy of Sciences, Beijing 100190, China}
\end{center}
\vspace{1em}

\section*{S1. Real-space charge and spin correlation functions}

To complement the momentum-space structure factors presented in the main text, we compute the real-space density-density connected correlation function $\langle n_i n_{i+r} \rangle_c$ and the spin-spin correlation function $\langle \mathbf{S}_i \cdot \mathbf{S}_{i+r} \rangle$ as functions of distance $r$ along the infinite direction of the cylinder, for typical finite bond dimensions $\chi$ and the $\chi \to \infty$ extrapolation.

\begin{figure}[htp]
\centering
\includegraphics[width=\textwidth]{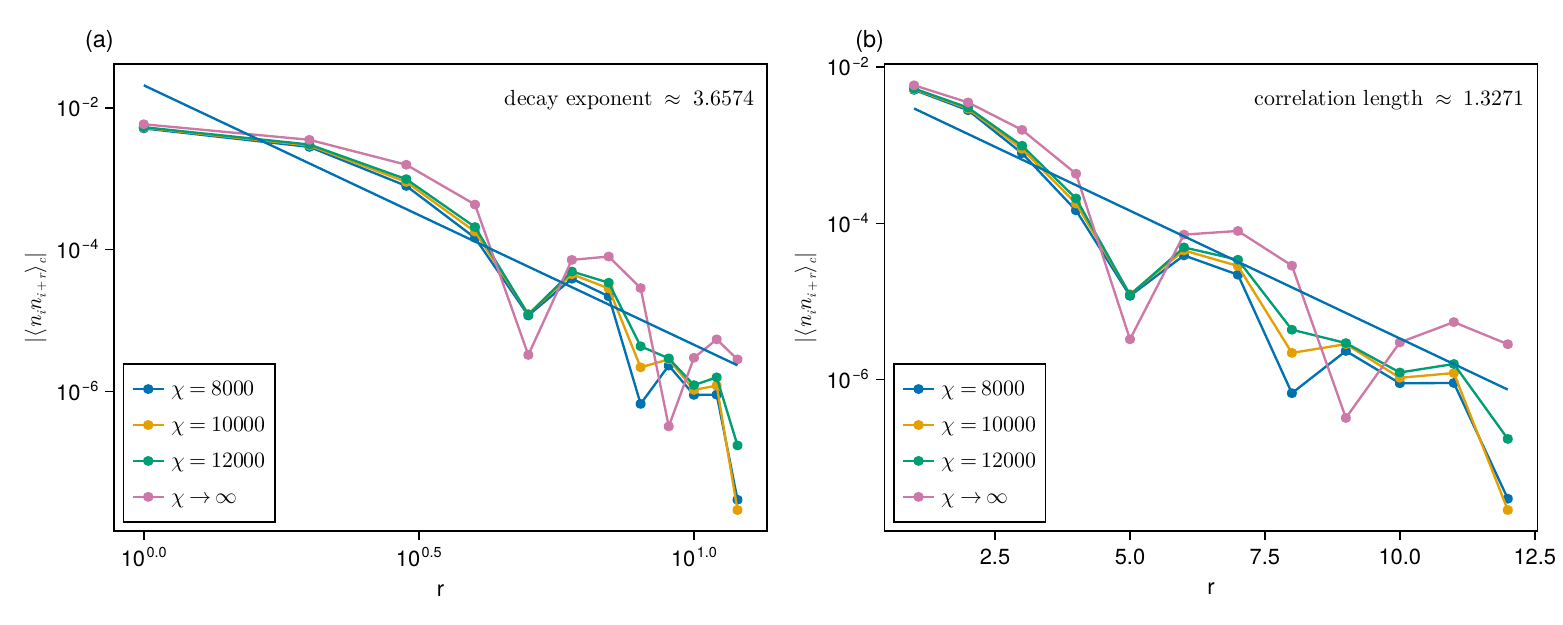}
\caption{Real-space density-density connected correlation function $|\langle n_i n_{i+r} \rangle_c|$, shown on (a) double-logarithmic and (b) semi-logarithmic scales. A power-law fit to the extrapolated curve gives a decay exponent $\approx 3.66$, while an exponential fit gives a correlation length $\xi \approx 1.33$ lattice constants. Both fits confirm the absence of long-range charge order.}
\label{fig:SM_density}
\end{figure}

\begin{figure}[htp]
\centering
\includegraphics[width=\textwidth]{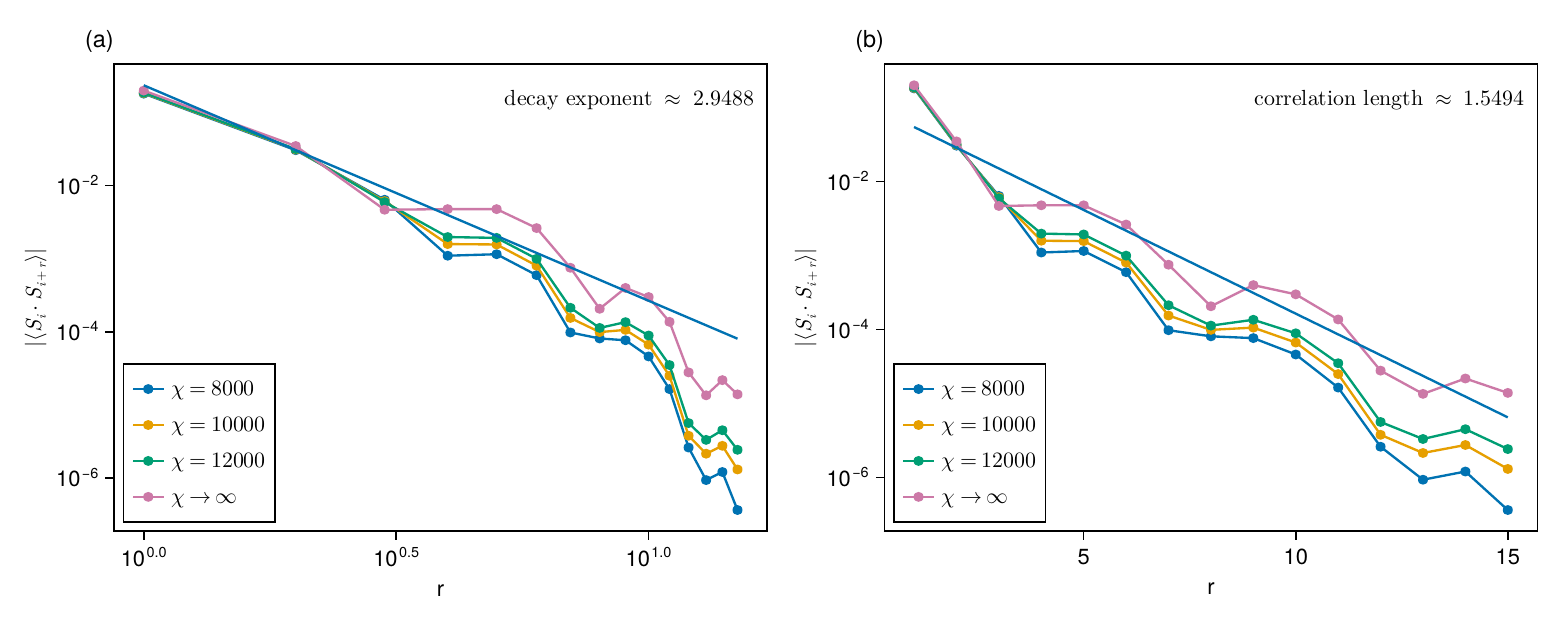}
\caption{Real-space spin-spin correlation function $| \langle \mathbf{S}_i \cdot \mathbf{S}_{i+r} \rangle |$, shown on (a) double-logarithmic and (b) semi-logarithmic scales. A power-law fit to the extrapolated curve gives a decay exponent $\approx 2.95$, while an exponential fit gives a correlation length $\xi \approx 1.55$ lattice constants. Both fits rule out long-range magnetic order.}
\label{fig:SM_spin}
\end{figure}

To characterize the decay without presuming its functional form, we fit the $\chi \to \infty$ extrapolated data to both a power law, $f(r) = A r^{-\alpha}$, and an exponential, $f(r) = B e^{-r/\xi}$, as shown in Figs.~\ref{fig:SM_density} and \ref{fig:SM_spin} on double-logarithmic and semi-logarithmic scales, respectively. For the charge sector, the power-law fit yields an exponent $\alpha \approx 3.66$, while the exponential fit yields a correlation length $\xi \approx 1.33$ lattice constants. For the spin sector, the corresponding values are $\alpha \approx 2.95$ and $\xi \approx 1.55$ lattice constants. The two fits describe the data comparably well over the accessible range of distances, and we therefore do not attempt to discriminate between them; what is robust is that, in either parametrization, both correlation functions decay extremely rapidly---with either a large power-law exponent or a correlation length of barely more than one lattice spacing. Such fast decays are consistent with short-range correlations only, confirming that neither the charge nor the spin sector exhibits long-range order, in agreement with the absence of sharp ordering peaks in static structure factors reported in the main text.

\section*{S2. Momentum distribution function}

\begin{figure}[htp]
\centering
\includegraphics[width=0.8\textwidth]{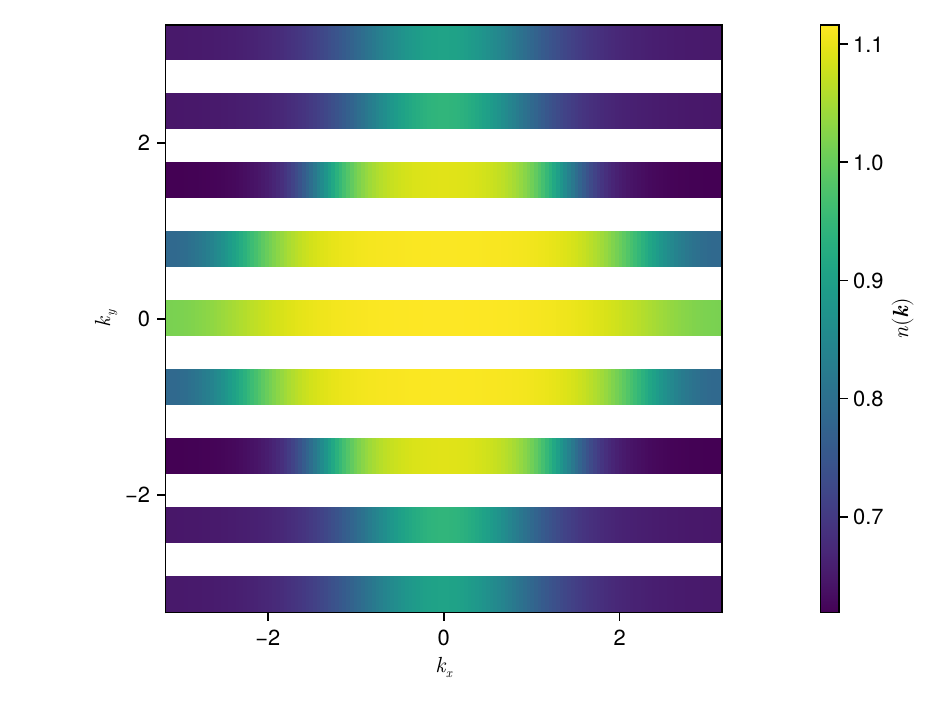}
\caption{Momentum distribution function $n(\mathbf{k})$ in the first Brillouin zone. The open, arc-like structure of the contours is qualitatively reminiscent of the Fermi arcs observed by ARPES in hole-doped cuprates, although the discrete $k_y$ cuts of the $L_y = 8$ cylinder limit the momentum resolution along $k_y$.}
\label{fig:SM_nk}
\end{figure}

The momentum-space electron distribution function, defined as
\begin{equation}
n(\mathbf{k}) = \sum_\sigma\langle c^\dagger_{\sigma\mathbf{k}} c_{\sigma\mathbf{k}} \rangle
= \frac{1}{N} \sum_{\mathbf{r}, \mathbf{r}', \sigma} \langle c^\dagger_{\sigma\mathbf{r}} c_{\sigma\mathbf{r}'} \rangle e^{i\mathbf{k} \cdot (\mathbf{r} - \mathbf{r}')},
\end{equation}
is shown in Fig.~\ref{fig:SM_nk}. The contour of steepest descent of $n(\mathbf{k})$ does not form a closed Fermi surface; instead, the spectral weight is organized into open, arc-like structures. We caution that the discrete $k_y$ cuts imposed by the finite cylinder circumference preclude a definitive identification of Fermi arcs; the observed structure is nevertheless qualitatively reminiscent of the arc-like spectral weight reported by ARPES in hole-doped cuprates~\cite{Damascelli2003}.

\section*{S3. Stripe order on the $L_y=6$ cylinder at $1/6$ doping}

\begin{figure}[htp]
\centering
\includegraphics[width=0.8\textwidth]{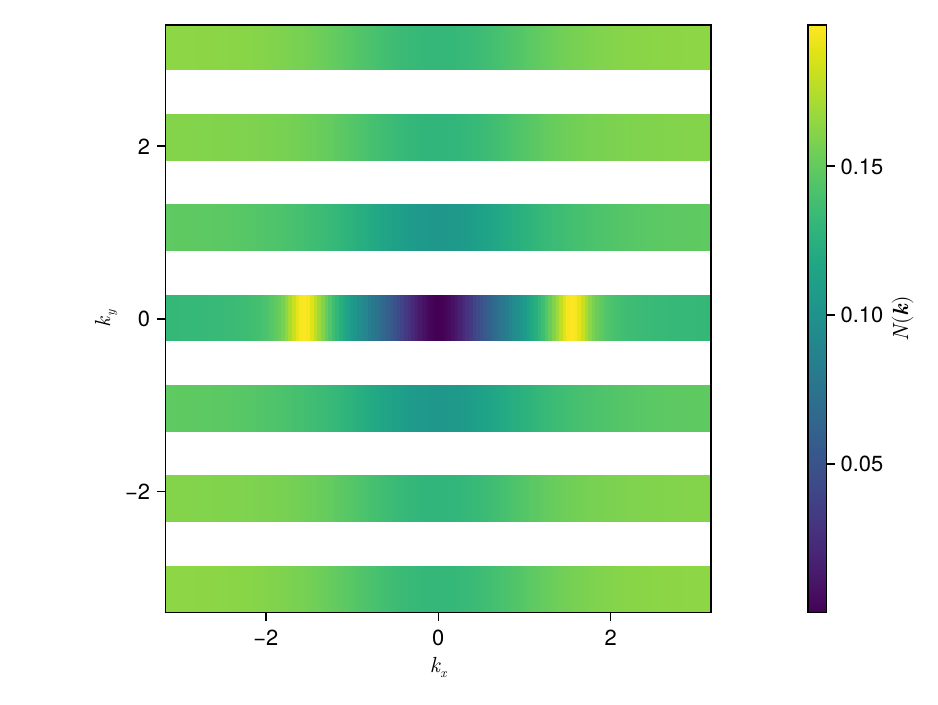}
\caption{Static charge structure factor $N(\mathbf{k})$ of the pure $t$-$J$ model ($J/t=0.4$) on the $L_y=6$ infinite cylinder at hole doping $\delta=1/6$, obtained with a $2\times6$ unit cell. Although the one-point density is uniform, $N(\mathbf{k})$ displays pronounced peaks at $(k_x,k_y)=(\pm\pi/2,0)$, corresponding to a period-4 charge stripe consistent with the period $4/(L_y\delta)$ found in earlier work~\cite{PhysRevLett.132.066002}.}
\label{fig:SM_6leg}
\end{figure}

To demonstrate that the absence of charge order on the $L_y=8$ cylinder at $1/8$ doping reported in the main text is not an artifact of the small period-2 unit cell, we apply the same VUMPS setup to the pure $t$-$J$ model ($J/t=0.4$) on an infinite cylinder of circumference $L_y=6$ at hole doping $\delta=1/6$, using a $2\times 6$ unit cell. The one-point density of the optimized ground state is spatially uniform, i.e., the density expectation value alone shows no symmetry breaking. The static charge structure factor $N(\mathbf{k})$, nevertheless, exhibits pronounced peaks at $(k_x,k_y)=(\pm\pi/2,0)$, as shown in Fig.~\ref{fig:SM_6leg}. These peaks correspond to a period-4 charge modulation along the cylinder axis, consistent with the stripe period $4/(L_y\delta)$ reported in a previous DMRG study of the hole-doped square-lattice $t$-$J$ model~\cite{PhysRevLett.132.066002}. 
This benchmark demonstrates that correlation-based diagnostics such as $N(\mathbf{k})$ remain sensitive to stripe modulations with periods exceeding the variational unit cell. It therefore supports our conclusion that the absence of pronounced charge-order peaks on the $L_y=8$ cylinder is not simply an artifact of the restricted unit-cell periodicity.

\end{document}